\documentclass[apjl]{emulateapj} 
\usepackage{natbib,amsmath,booktabs,apjfonts,threeparttable,multirow}
\newcommand{\snr}{RCW~86}

\newcommand{\xmm}{{\it XMM--Newton}}
\newcommand{\chandra}{{\it Chandra}}

\newcommand\suzaku{{\it Suzaku}}

\shortauthors{Castro et al.}
\shorttitle{CHANDRA VIEW OF THE NONTHERMAL EMISSION IN THE NW OF SNR RCW~86}
\slugcomment{Submitted to ApJ}

\begin{document}
\title{A Chandra view of Nonthermal emission in the Northwestern Region of Supernova Remnant RCW~86: particle acceleration and magnetic fields}
\author{Daniel Castro\altaffilmark{1}, Laura A. Lopez\altaffilmark{1}, Patrick O. Slane\altaffilmark{2}, Hiroya Yamaguchi\altaffilmark{2,3,4}, Enrico Ramirez--Ruiz\altaffilmark{5}, Enectali Figueroa--Feliciano\altaffilmark{1}}

\altaffiltext{1}{MIT-Kavli Center for Astrophysics and Space Research, 77 Massachusetts Avenue, Cambridge, MA, 02139, USA}
\altaffiltext{2}{Harvard-Smithsonian Center for Astrophysics, 60 Garden Street, Cambridge, MA 02138, USA}
\altaffiltext{3}{NASA Goddard Space Flight Center, Code 662, Greenbelt, MD 20771, USA}
\altaffiltext{4}{Department of Astronomy, University of Maryland, College Park, MD 20742, USA}
\altaffiltext{5}{Department of Astronomy and Astrophysics, University of California Santa Cruz, 1156 High Street, Santa Cruz, CA 95060, USA}

\begin{abstract}

The shocks of supernova remnants (SNRs) are believed to accelerate particles to cosmic ray (CR) energies. The amplification of the magnetic field due to CRs propagating in the shock region is expected to have an impact on both the emission from the accelerated particle population, as well as the acceleration process itself. 
Using a 95 ks observation with the Advanced CCD Imaging Spectrometer (ACIS) onboard the \chandra\ {\it X-ray Observatory}, we map and characterize the synchrotron emitting material in the northwestern region of \snr. We model spectra from several different regions, filamentary and diffuse alike, where emission appears dominated by synchrotron radiation. The fine spatial resolution of Chandra allows us to obtain accurate emission profiles across 3 different non-thermal rims in this region. The narrow width ($l \approx10''-30''$) of these filaments constrains the minimum magnetic field strength at the post-shock region to be approximately 80 $\mu$G. 

\end{abstract}


\section{Introduction}

\noindent Non-thermal X-ray emission has been detected from several young shell-type supernova remnants (SNRs), including SN~1006 \citep{koyama_1995}, RX J1713.7--3946 \citep{koyama_1997}, and Vela Jr. \citep{aschenbach_1998}. These X-rays are believed to be synchrotron radiation from electrons accelerated to TeV energies at the shocks, interacting with the compressed, and possibly amplified, local magnetic field. Observations of $\gamma$-ray emission from several SNRs in the TeV range confirm that particles are being accelerated to energies approaching the knee of the cosmic ray spectrum in these remnants \citep[e.g.][]{aharonian_2004,aharonian_2005, naumann_2008}. However, while it is broadly believed that diffusive shock acceleration (DSA) in SNRs produces the bulk of cosmic rays below 10$^{15}$ eV, we still lack  a detailed understanding of the acceleration process and its effects on the the system, such as magnetic field amplification (MFA) and maximum particle energy. 

Since the amplification of the magnetic field due to the cosmic-ray acceleration process is expected to have an impact on both the emission from the accelerated particle population as well as the acceleration process itself, it has become a crucial area of research \citep[e.g.][]{vladimirov_2006}. Observations of heliospheric shocks more than three decades ago led to suggestions that strong shocks, such as those in SNRs, could amplify the ambient magnetic field \citep{chevalier_1977}. Multiwavelength observations and models of young SNRs also suggest that the magnetic fields at the forward shock are much larger than expected from simple compression of the ambient field of the ISM. This is evidenced through several different observed effects, including the broadband spectrum of the synchrotron emission from radio to X-rays of several SNRs \citep[e.g.,][]{volk_2005} and the rapid variability of bright knots of non-thermal emission in some remnants \citep[e.g.,][]{uchiyama_2007}. The inferred magnetic fields in SNR Cassiopeia A are $B\sim0.5$ mG, and similar values ($B\gtrsim0.1$ mG) have been estimated from observations of Tycho, Kepler, SN 1006, and G349.7--0.5 \citep{reynolds_1992,volk_2005,uchiyama_2007}. Given that the ISM magnetic field is $B_{\text{ISM}}\sim3-10\,\mu$G, the strengths of the fields inferred from observations of SNRs imply amplification factors of order 10--100, likely an effect of DSA at the shocks of these remnants. \citet{bell_1978} suggests that magnetic field amplification (MFA) is the result of non-resonant cosmic ray instabilities in the shock precursor. MFA is a crucial element in the DSA process since the turbulent field created by CRs is responsible for the scattering of particles in the shock, leading to their acceleration to CR energies. Several authors \citep[e.g., ][]{vink_2003} have shown that the magnetic field strength of the post-shock region is closely linked to the width of X-ray synchrotron filaments in SNRs. 

 \begin{figure*}
\includegraphics[width=0.495\textwidth]{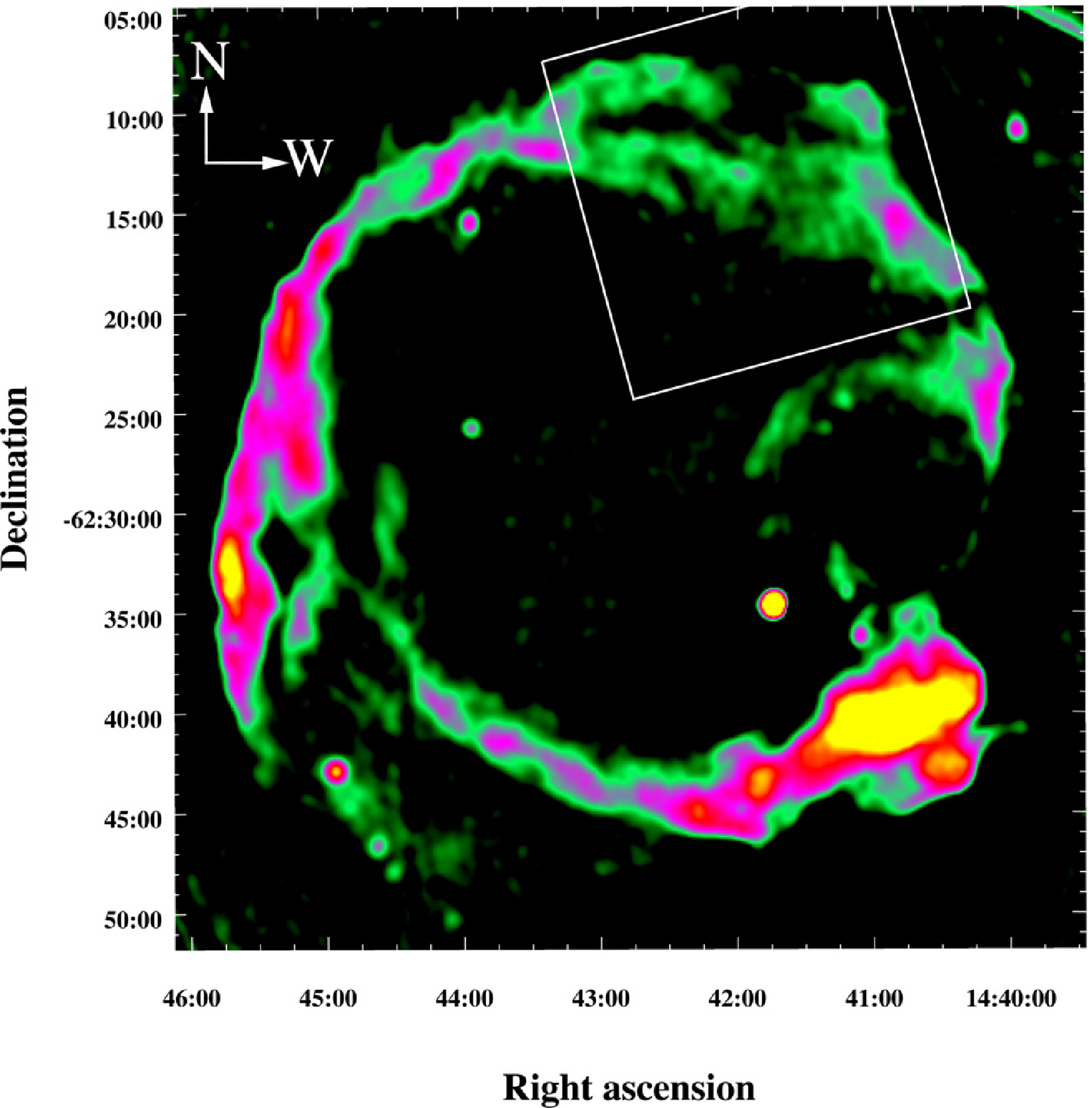}
\includegraphics[width=0.495\textwidth]{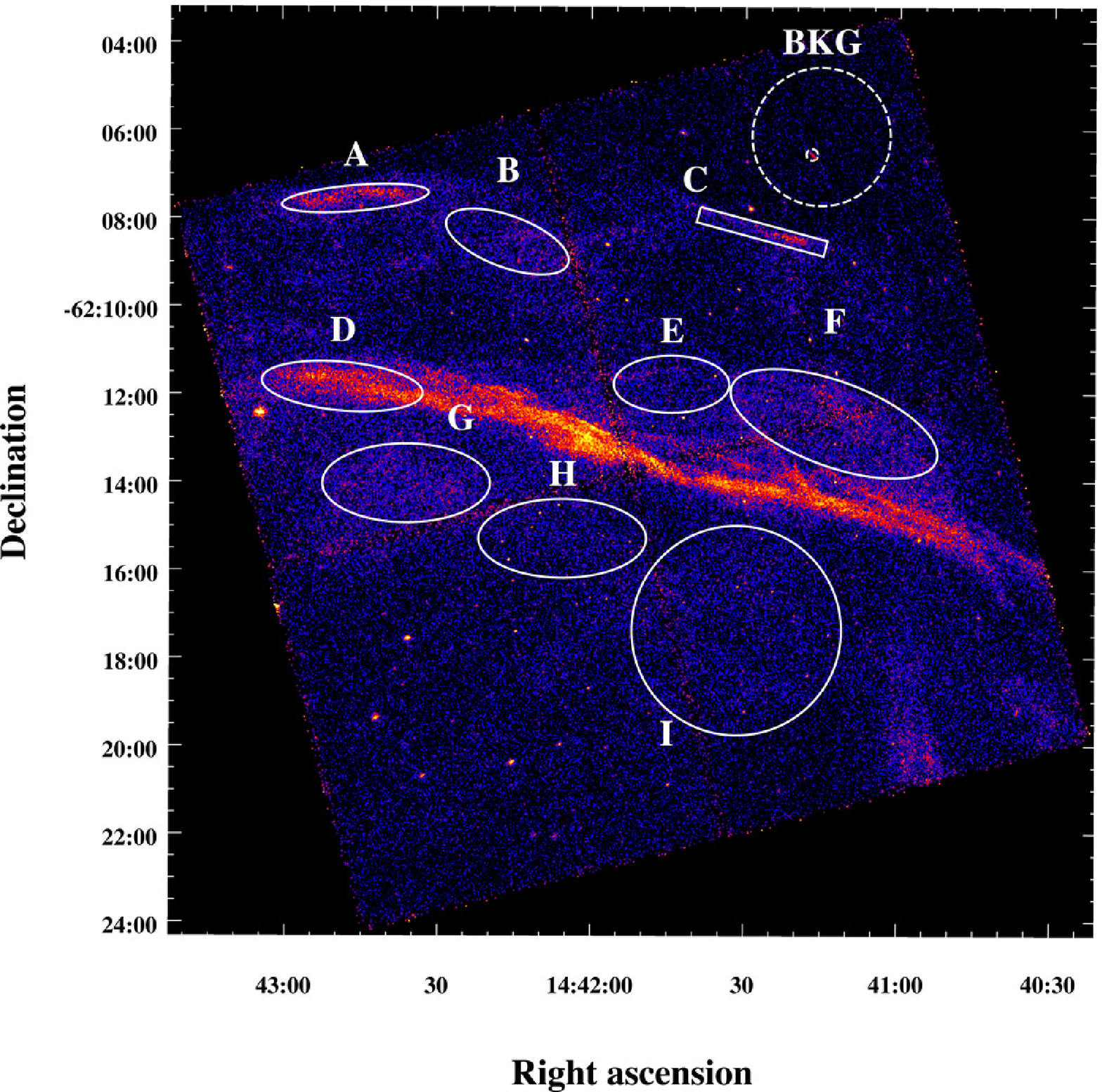}\\
\includegraphics[width=0.495\textwidth]{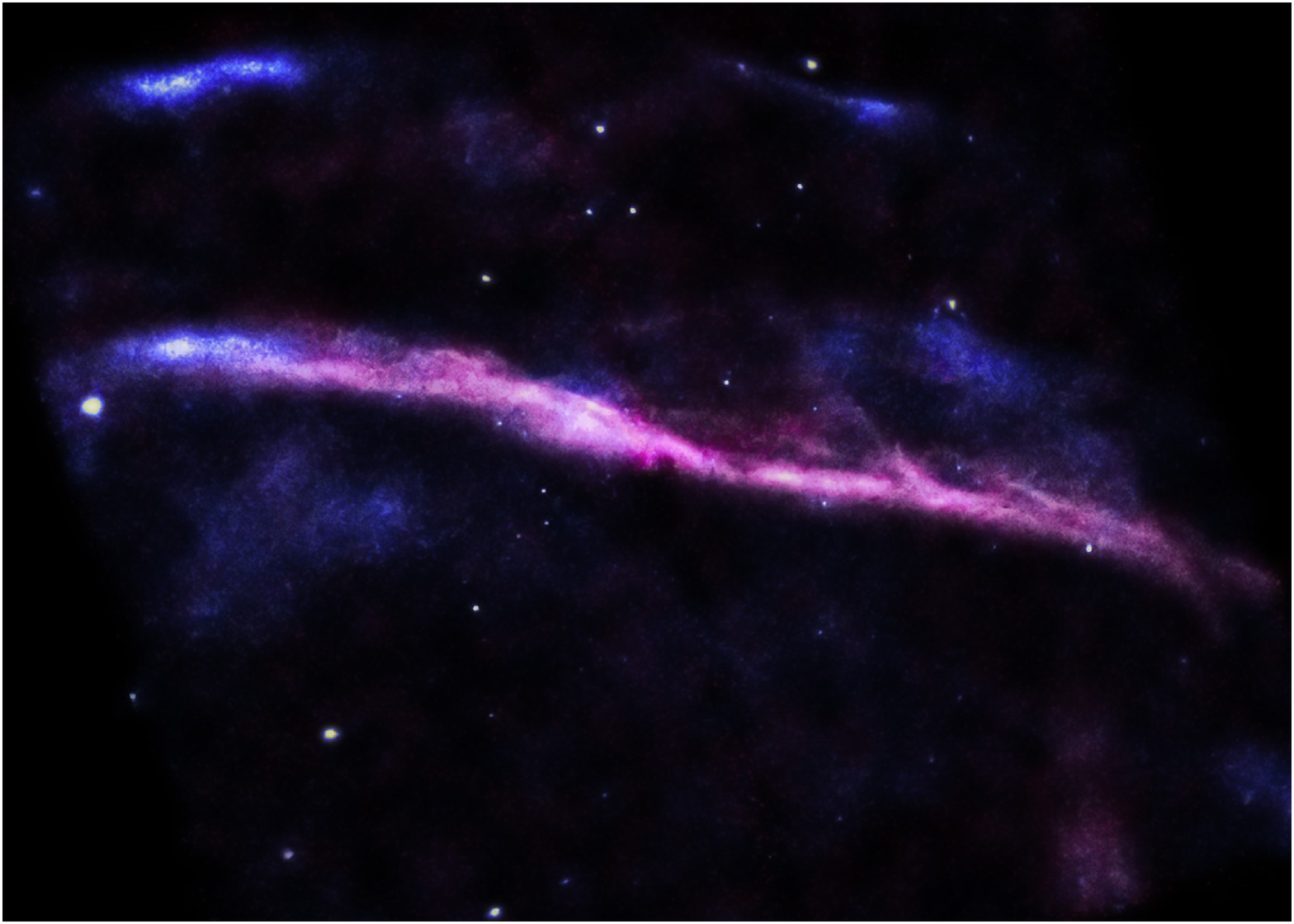}
\includegraphics[width=0.495\textwidth]{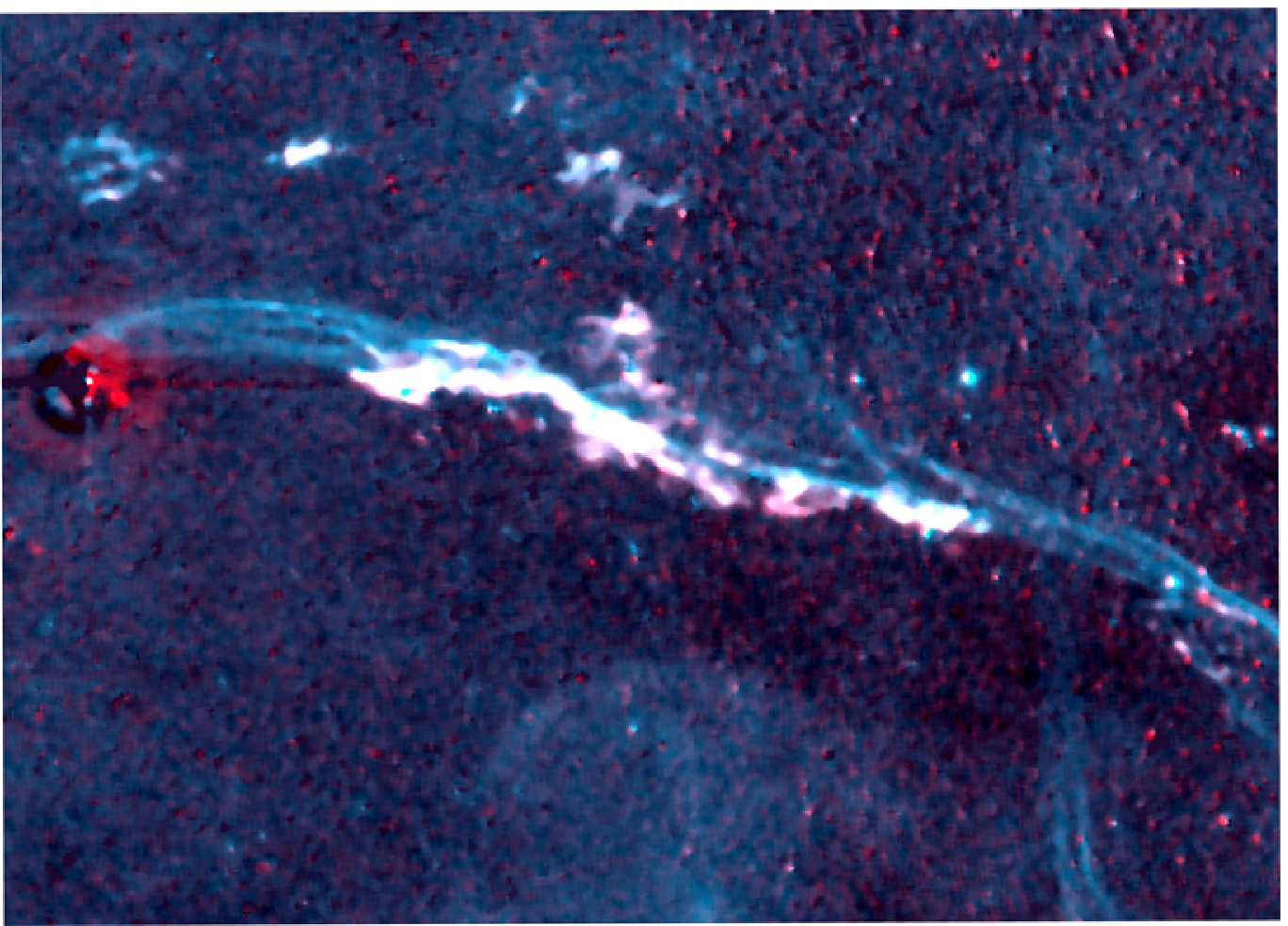}
\caption{\footnotesize \emph{Top left panel:} Radio image of the SNR \snr\ from observations with the Molonglo Synthesis Telescope at 843 MHz, with square root intensity scale and spanning 0 to 0.2 Jy/beam \citep{whiteoak_1996}. The morphology of \snr\ is clearly shell-type, yet it does not appear spherical and suggests that the shock has interacted with regions of different density in different areas. The white square indicates the region observed with \chandra\ and discussed in this paper. \emph{Top right panel:} Exposure corrected image of the NW of \snr\ in the 0.5--7.0 keV energy range, with pixel size 0$''$.492, smoothed with a gaussian with kernel $\sigma=3$ pixels, and using a logarithmic intensity scale. The X-ray emission defines a main emission rim much more clearly than the radio emission in this region, and two small arcs are observed $\sim5'$ north of its position. Overlaid in white are the 9 regions used for spectral analysis as well as the background region selected. \emph{Bottom left panel:}  Two-color image created by combining soft X-ray emission (0.3-0.75 keV, in magenta) and the hard band (1.5--7 keV in blue).  The thin arcs ahead of the shock appear to be dominated by hard X-rays, as is the eastern part of the main rim. Image processing credit: This image was produced in collaboration with NASA/CXC/SAO/J.DePasquale. \emph{Bottom right panel}: Two-color optical image of the region (S~{\scshape ii} in red, H$\alpha$ in cyan), created using observations with the 0.9m Curtis/Schmidt telescope at the Cerro Tololo Inter-American Observatory (CTIO) by \citet{smith_1997}. }
\label{fig:broad}
\end{figure*}

RCW~86 (G315.4--2.3) is a large ($\sim 40'$ across) Galactic SNR possibly associated with the historical supernova explosion SN 185 \citep[][and references therein]{vink_2006}. It displays a shell-type morphology in radio \citep{kes_1987}, optical \citep{smith_1997}, and X-rays \citep{pisarski_1984}. The radio observation with the Molonglo Synthesis Telescope at 843 MHz, shown in Figure \ref{fig:broad} ({\it top left panel}), illustrates the broad morphology of RCW~86 \citep{whiteoak_1996}. Optical studies have derived kinematic distances to RCW~86 of 2.3$\pm0.2$ and 2.8$\pm0.4$ kpc \citep[][respectively]{sollerman_2003,rosado_1996}; hereafter, we adopt $d=2.5$ kpc. This value is also consistent with the distance to a group of OB stars found in the region and possibly related to the progenitor system \citep{westerlund_1969}. The entire extent of the SNR has been covered by \xmm\ observations, which reveal both thermal and non-thermal emission \citep{vink_2006}. Extended TeV $\gamma$-ray emission has recently been detected in the north and south regions of the SNR with the HESS Cherenkov Telescope \citep{aharonian_2009}. It is hence clear, both from the non-thermal X-ray emission observations and the $\gamma$-ray detections, that RCW~86 accelerates particles up to cosmic-ray energies. 

\chandra\ observations taken along the NE and SW rims of RCW~86 resolved the spatial distribution of the thermal and non-thermal emission regions \citep{rho_2002}. In typical shell-type SNRs such as Tycho, Cas A and SN 1006, non-thermal X-ray emission is observed as thin filamentary structures close to the forward shock. The \chandra\ study of the NE rim of \snr\ shows that indeed the X-ray emission is dominated by non-thermal emission located at the blast wave \citep{vink_2006}. In contrast, the non-thermal emission to the SW in RCW~86 is much broader and it is not confined to the blast wave region \citep{rho_2002}. Previous studies of the NW region with \xmm\ and \suzaku\ have revealed it to be highly complex area spectrally and morphologically \citep[][respectively]{williams_2011, yamaguchi_2011}. These analyses found that the X-ray bright rim and some diffuse material ahead of it show signatures of both ejecta and non-thermal emission. 

\begin{figure*}
\begin{center}
\includegraphics[width=0.32\textwidth]{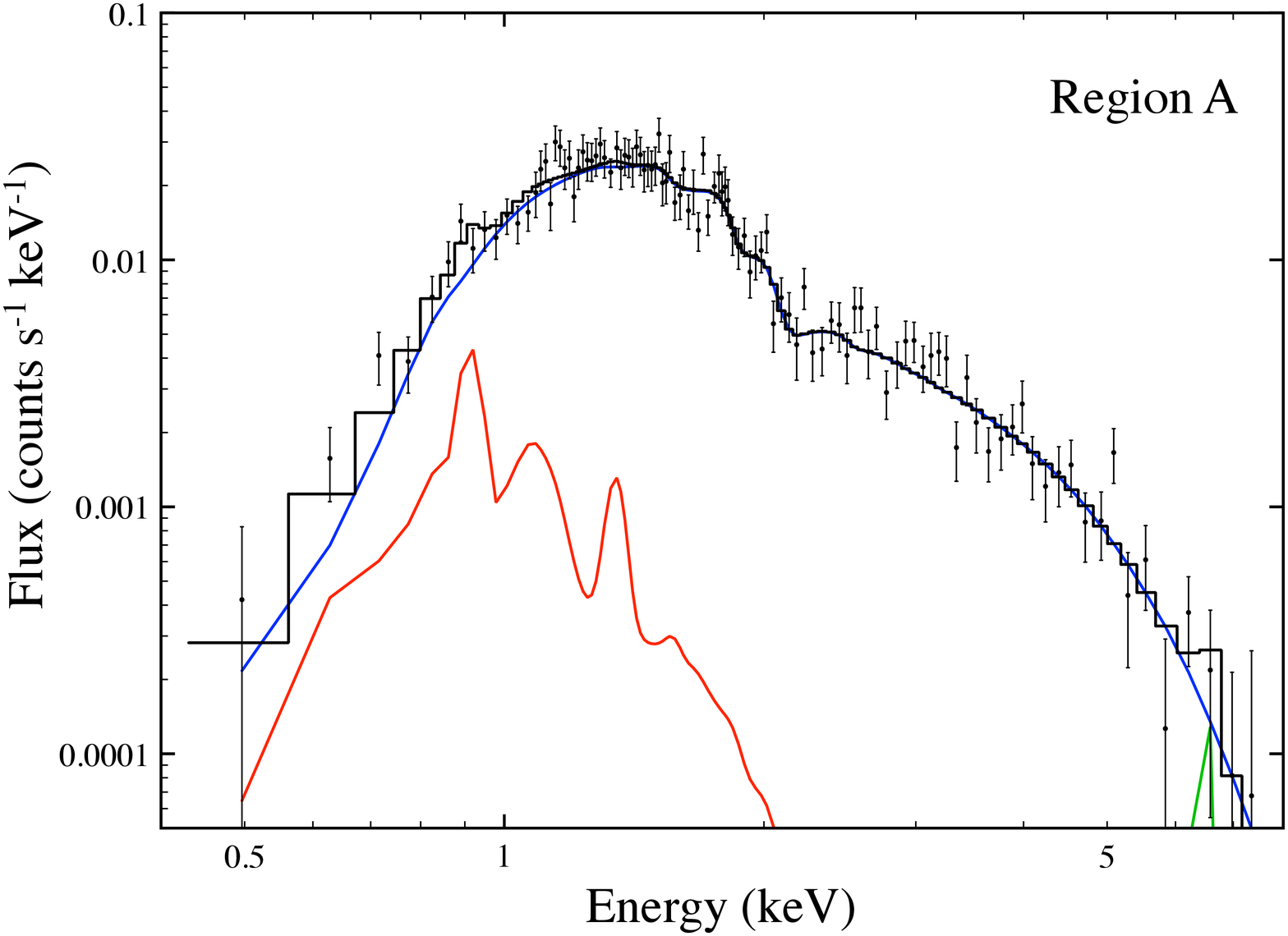}
\includegraphics[width=0.32\textwidth]{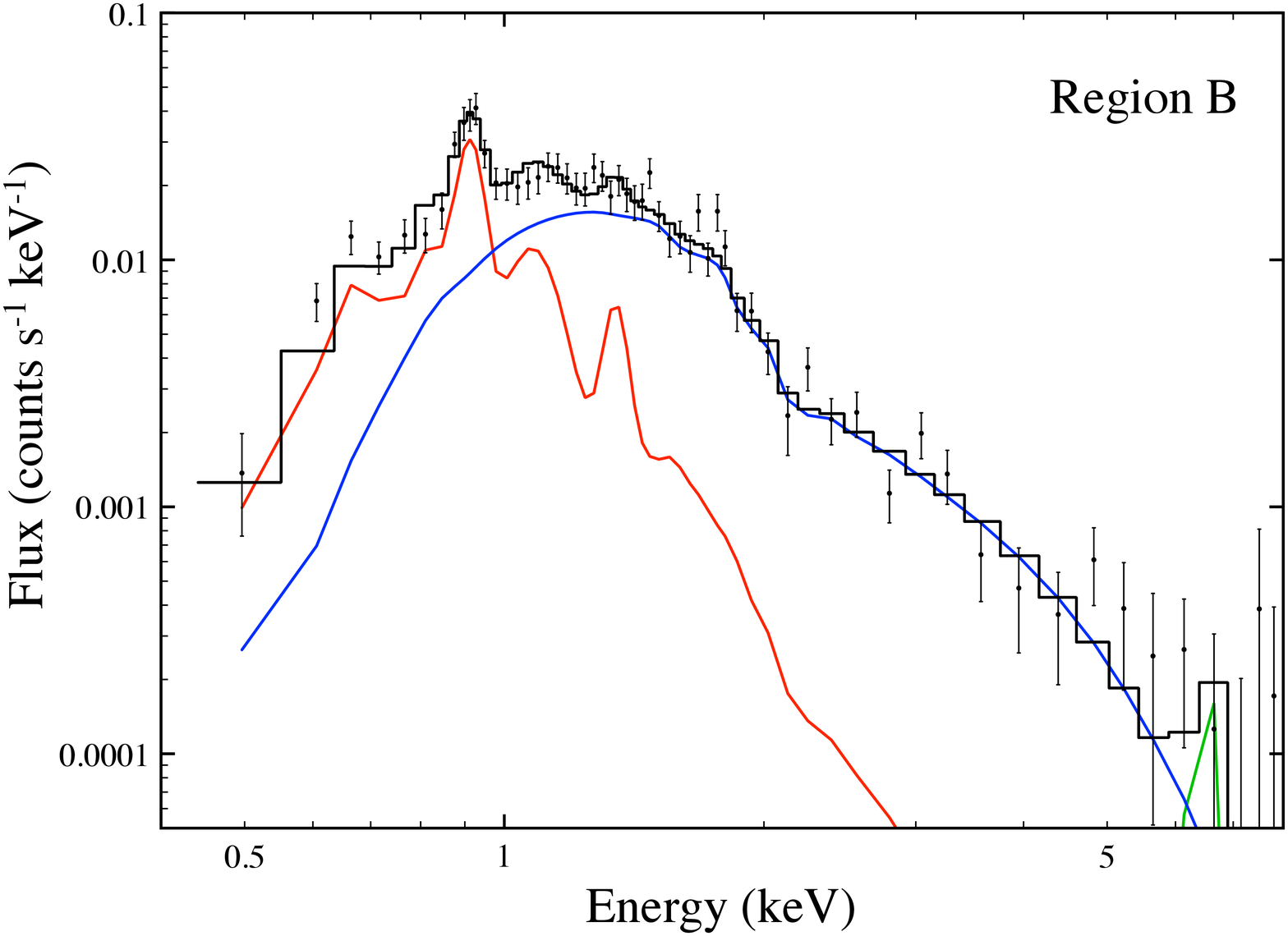}
\includegraphics[width=0.32\textwidth]{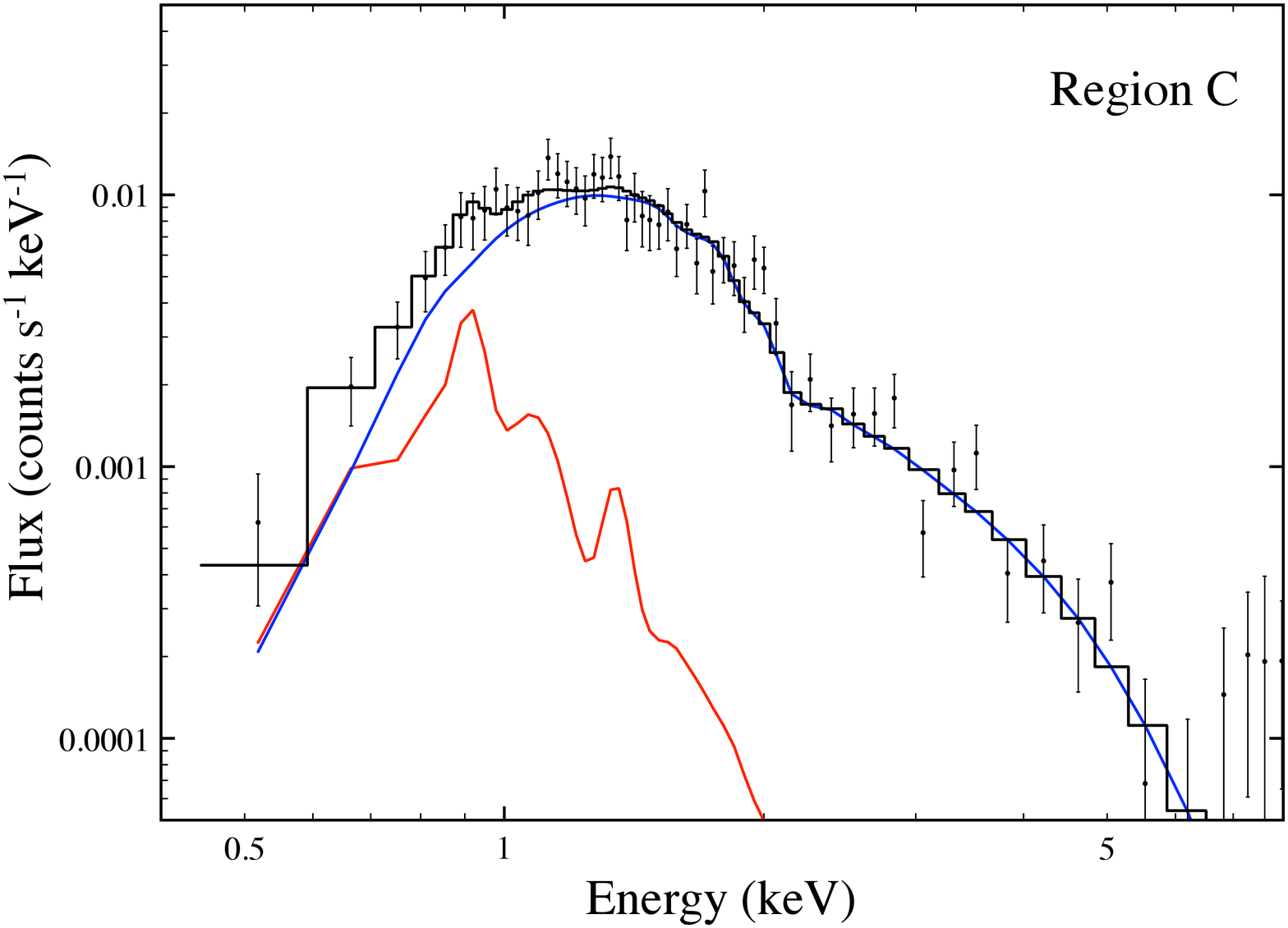}\\
\includegraphics[width=0.32\textwidth]{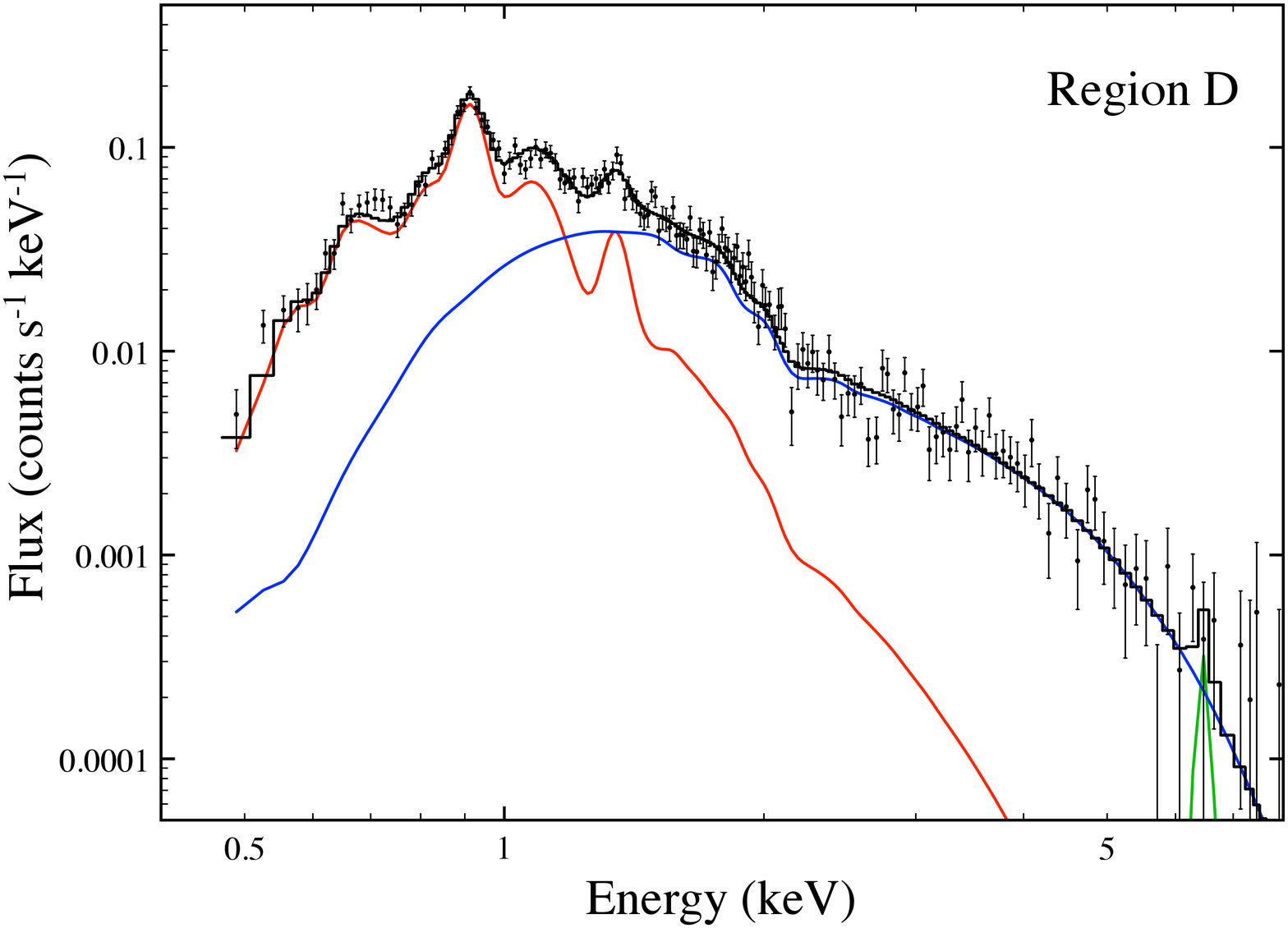}
\includegraphics[width=0.32\textwidth]{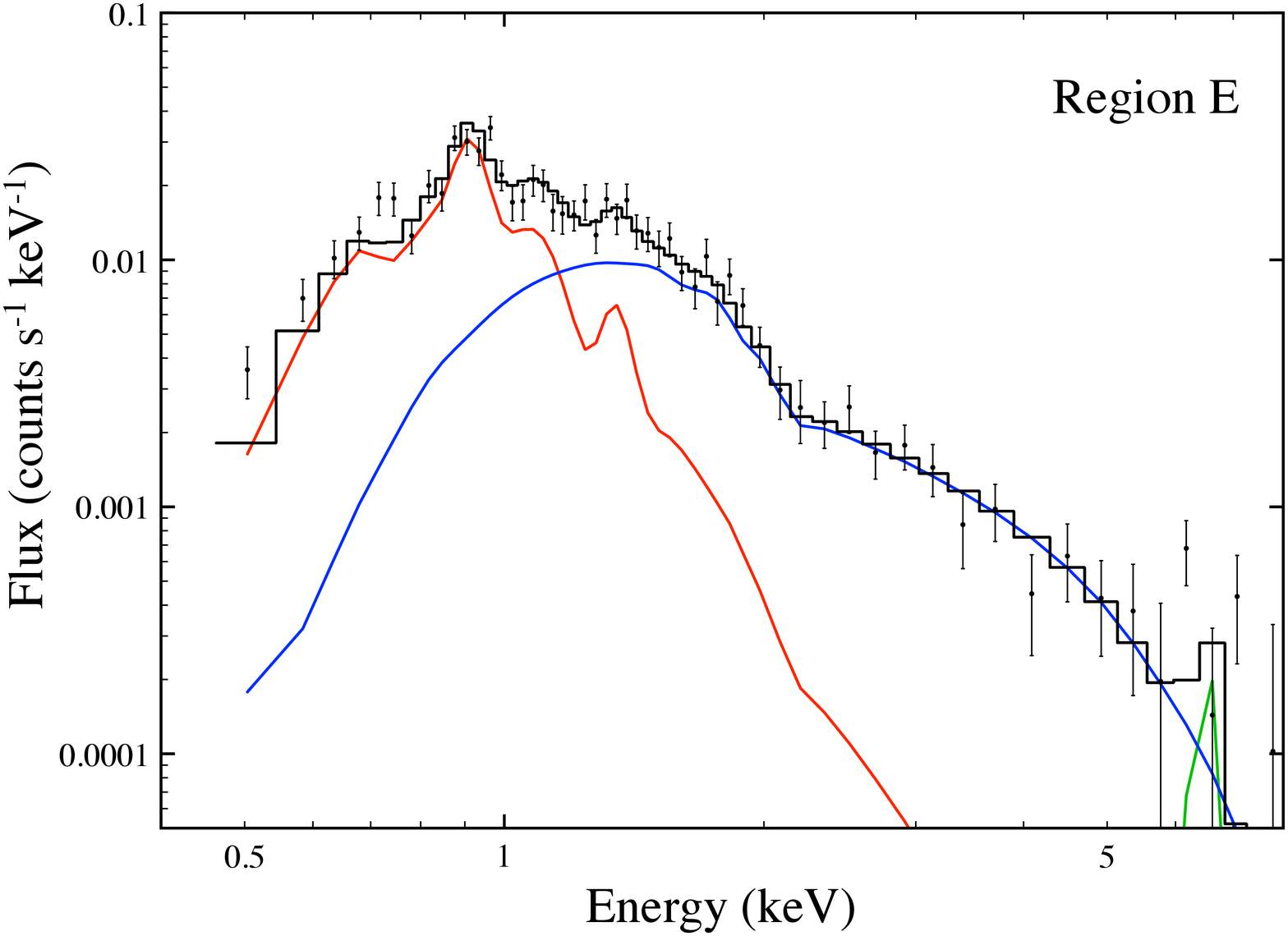}
\includegraphics[width=0.32\textwidth]{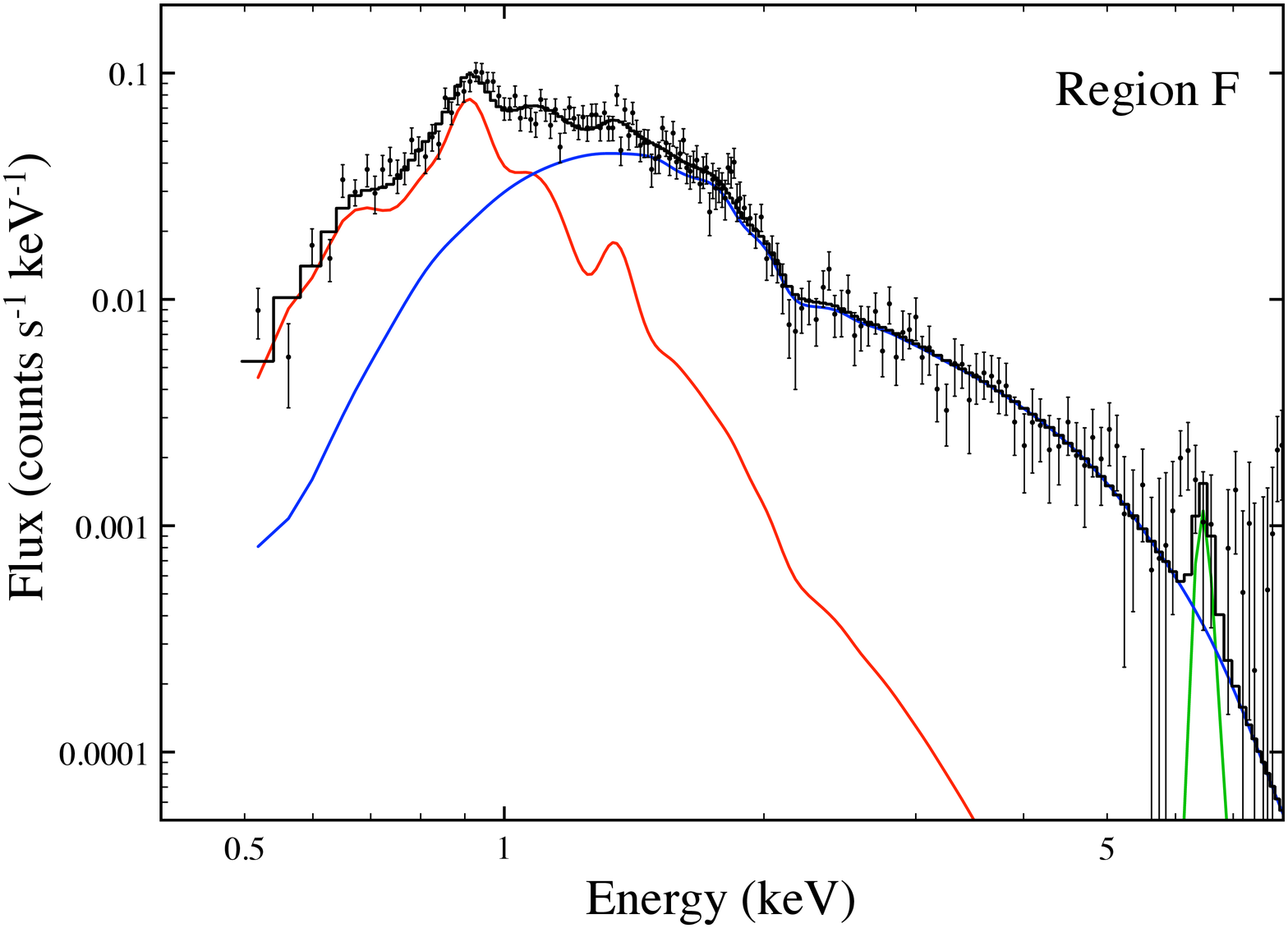}\\
\includegraphics[width=0.32\textwidth]{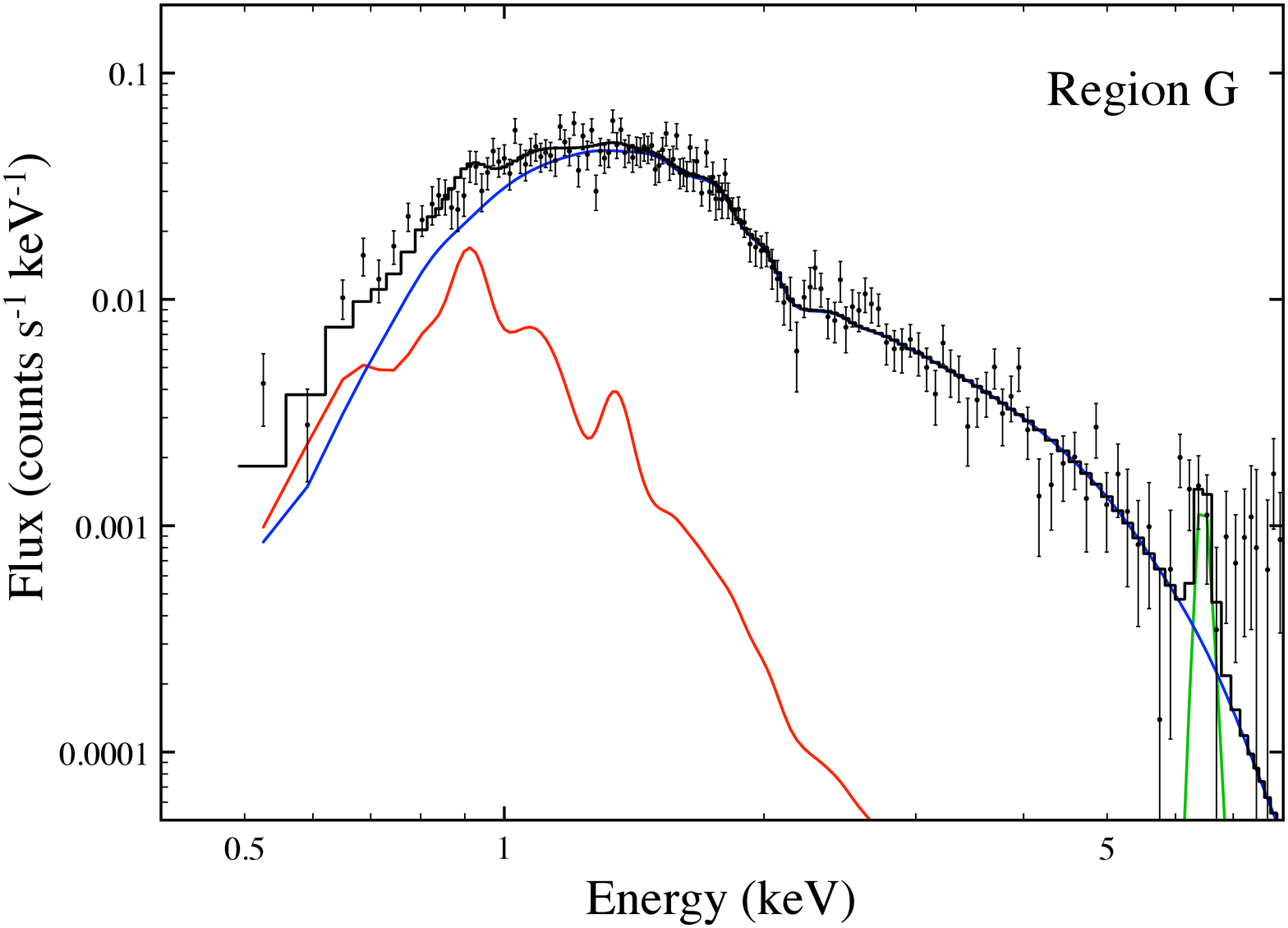}
\includegraphics[width=0.32\textwidth]{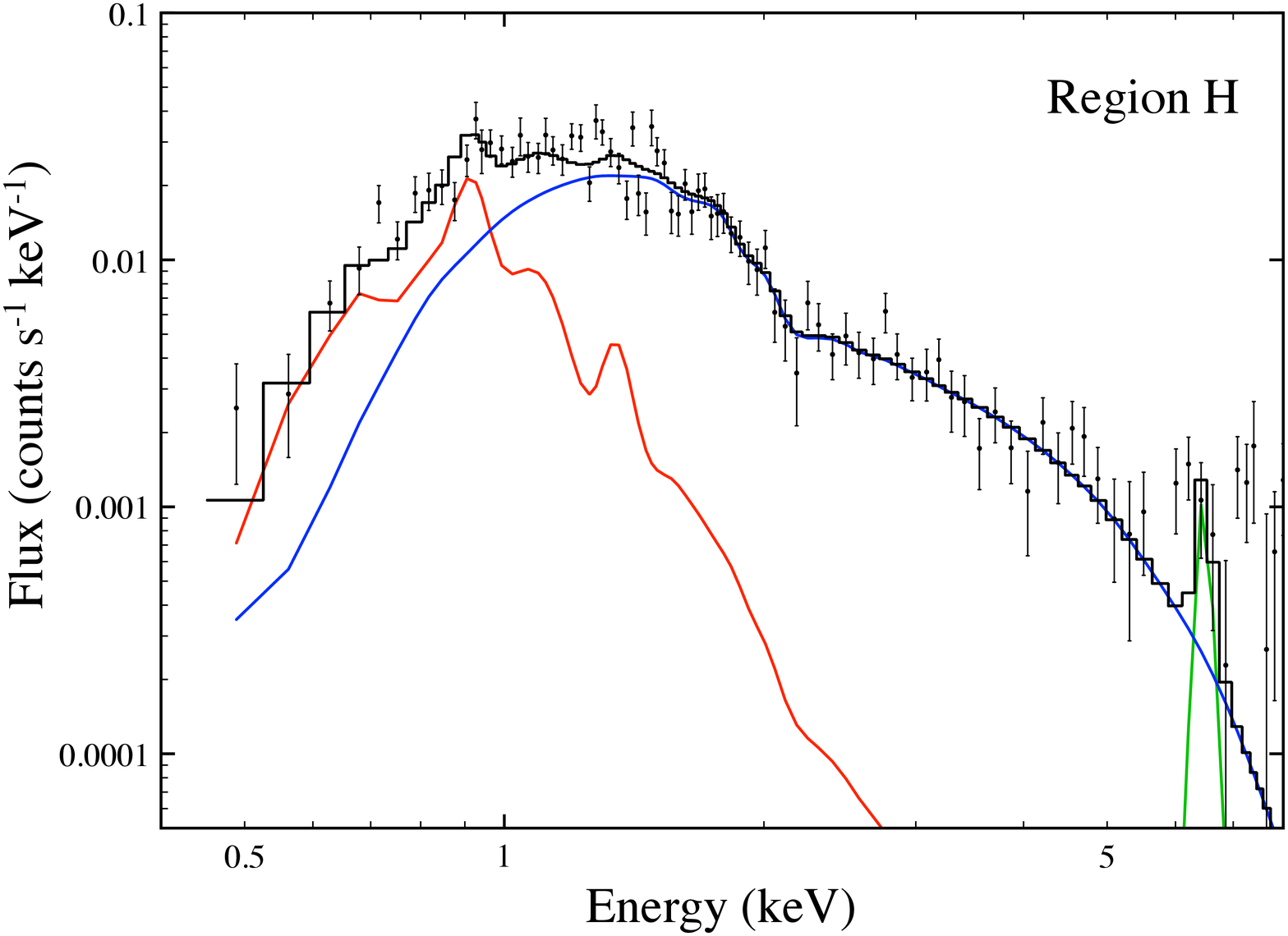}
\includegraphics[width=0.32\textwidth]{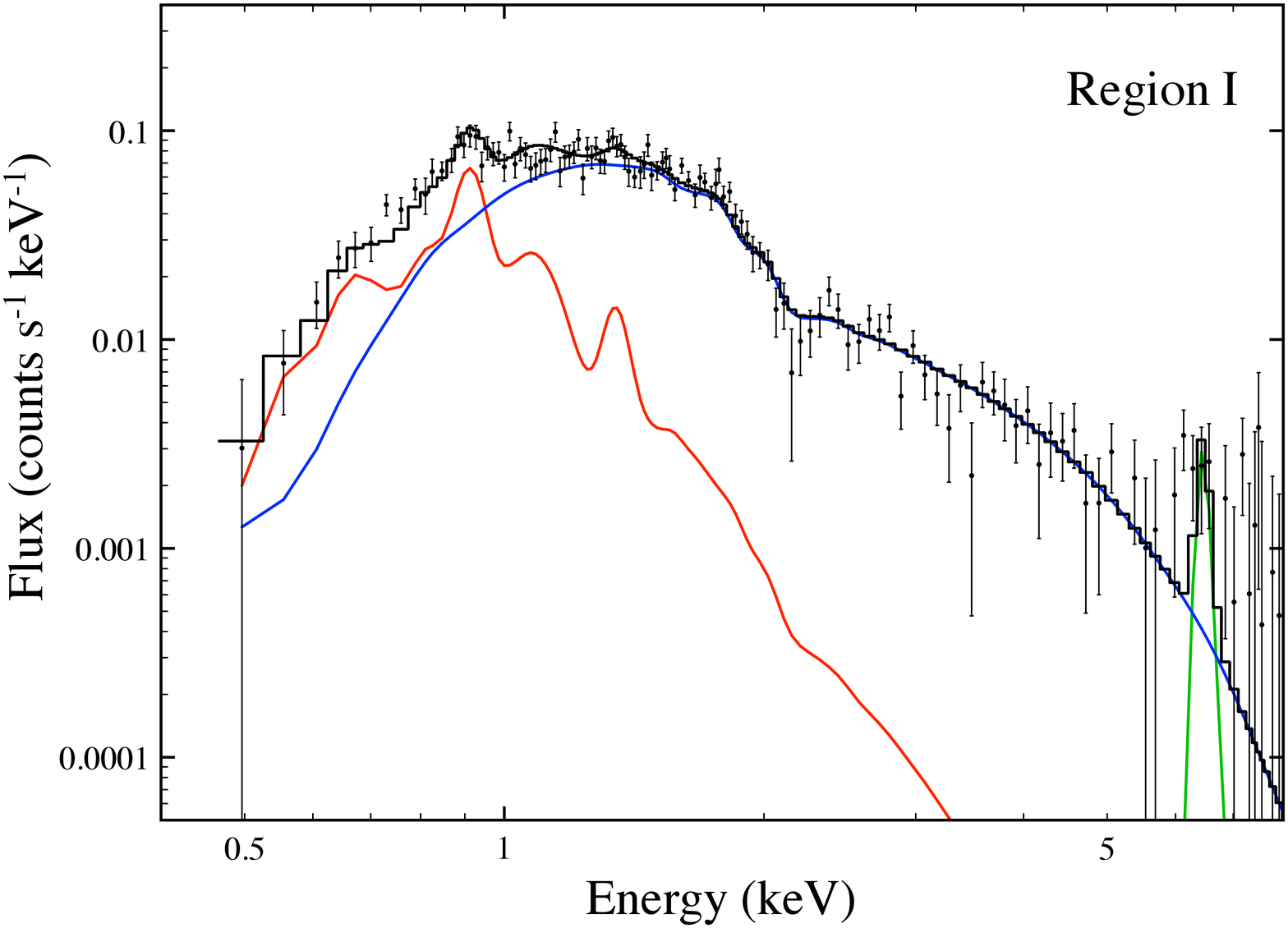}\\
\end{center}
\caption{\footnotesize Background subtracted spectra extracted from 9 regions in NW of \snr\ (shown in Figure \ref{fig:broad} \emph{top right panel}). All have been fit using models combining absorbed NEI and non thermal emission as outlined in Section 2. Black histograms show the total best fit model (with parameters shown in Table \ref{tab:spectral}), blue curves show non-thermal emission, red lines show NEI emission, and the green component is a gaussian model representing the Fe-K line (6.45 keV). All spectra are dominated by non-thermal emission.}
\label{fig:spectra}
\end{figure*}

In this work, we map and characterize the synchrotron emitting material in the NW of \snr, and hence constrain the post-shock magnetic field and the shape of the non-thermal emission spectra in this region. For this study we use a 95 ks observation with the Advanced CCD Imaging Spectrometer (ACIS) onboard the \chandra\ {\it X-ray Observatory}. We take advantage of the fine spatial resolution of \chandra\ to obtain emission profiles across 3 different non-thermal rims in the region and derive the minimum magnetic field magnitudes for such filament widths. Additionally, we model spectra from several different regions, filamentary and diffuse alike, where emission appears dominated by synchrotron radiation. In Section 2 we describe the observational data and how they have been analyzed to obtain emission profiles and spectra, and we discuss models used to fit both. Finally, in Section 3 we discuss how the observations are interpreted to derive magnetic field magnitudes and to constrain the shape of the accelerated electron distributions.

\begin{center}
\begin{deluxetable*}{cccccccccccccccccc}
\tabletypesize{\footnotesize}
\tablecaption{\footnotesize{Results of Spectral Fitting}\label{tab:spectral}}
\startdata
\toprule
\noalign{\smallskip}
&&\multicolumn{7}{c}{SRCUT}&&&\multicolumn{7}{c}{Power-Law}\\ 
\noalign{\smallskip}
\cline{3-9} \cline{12-18}
\noalign{\smallskip}
\multirow{2}{*}{Region}&&$h\nu_{\text{roll-off}}$&&$F_{\text{srcut}}$\tablenotemark{a}&&$F_{\text{nei}}$\tablenotemark{a}&&\multirow{2}{*}{$\chi^2$/dof}&&&\multirow{2}{*}{$\Gamma$}&&$F_{\text{powerlaw}}$\tablenotemark{a}&&$F_{\text{nei}}$\tablenotemark{a}&&\multirow{2}{*}{$\chi^2$/dof}\\
\noalign{\smallskip}
&&(keV)&\multicolumn{4}{c}{($10^{-13}$ erg cm$^{-2}$ s$^{-1}$)} &&&&&&\multicolumn{4}{c}{($10^{-13}$ erg cm$^{-2}$ s$^{-1}$)}&&\\
\noalign{\smallskip}
\midrule
A............&	&	$0.37^{+0.05}_{-0.06}$	&	&	1.93&	&	$0.004$&	&	108.2/106	&	&	&	$2.65^{+0.06}_{-0.06}$	&	&	1.92&	&	0.003&	&	110.3/106\\
B............&	&	$0.12^{+0.03}_{-0.02}$	&	&	0.68&	&	$0.023$&	&	57.4/55	&	&	&	$3.1^{+0.1}_{-0.1}$		&	&	0.71&	&	0.021&	&	58.6/55\\
C............&	&	$0.17^{+0.06}_{-0.04}$	&	&	0.54&	&	$0.003$&	&	47.1/54	&	&	&	$2.9^{+0.1}_{-0.1}$		&	&	0.56&	&	0.003&	&	47.1/54\\
D............&	&	$0.28^{+0.03}_{-0.05}$	&	&	2.39&	&	$0.147$&	&	190.8/166	&	&	&	$2.78^{+0.06}_{-0.06}$	&	&	2.44&	&	0.144&	&	190.5/166\\
E............&	&	$0.6^{+0.3}_{-0.2}$		&	&	0.72&	&	$0.026$&	&	63.8/53	&	&	&	$2.5^{+0.1}_{-0.1}$		&	&	0.74&	&	0.025&	&	64.4/53\\
F............&	&	$0.46^{+0.06}_{-0.09}$	&	&	3.32&	&	$0.080$&	&	173.9/162	&	&	&	$2.59^{+0.06}_{-0.06}$	&	&	3.40&	&	0.077&	&	171.5/162\\
G............&	&	$0.32^{+0.06}_{-0.04}$	&	&	3.03&	&	$0.016$&	&	167.7/143	&	&	&	$2.68^{+0.05}_{-0.05}$	&	&	3.05&	&	0.012&	&	167.8/143\\
H............&	&	$0.8^{+0.2}_{-0.2}$		&	&	1.76&	&	$0.018$&	&	134.9/87	&	&	&	$2.43^{+0.08}_{-0.08}$	&	&	1.76&	&	0.016&	&	131.1/87\\
I.............&	&	$0.27^{+0.05}_{-0.04}$	&	&	4.03&	&	$0.052$&	&	149.2/125	&	&	&	$2.75^{+0.06}_{-0.06}$	&	&	4.14&	&	0.047&	&	150.4/125
\enddata 
\noindent \tablecomments{\noindent Absorption model and solar abundance values obtained from \citet{wilms_2000}. The absorbing column density is set to $N_{\text{H}}=0.6\times10^{22}$ atoms cm$^{-2}$. The electron temperature and ionization timescale of the {\scshape nei} component were fixed at $kT_{\text{e}}=0.56$ keV and $n_{\text{e}}t=10^{10}$s cm$^{-3}$ respectively.
}
\tablenotetext{a}{Unabsorbed fluxes in the 2--6 keV energy range.}

\end{deluxetable*}
\end{center}

\section{Observations and Analysis}

\noindent The northwestern rim of \snr\ was observed with \chandra\ ACIS-I for 95 ks on 3 and 11 February 2013 (ObsIDs 14890, 15608 and 15609) in the {\scshape timed} exposure {\scshape vfaint} mode. All data analyses were performed using the \emph {Chandra} Interactive Analysis of Observations ({\scshape ciao}) software package version 4.5 \citep{ciao_2006}. 

In Figure \ref{fig:broad} (\emph{top right}) we show an exposure corrected image of the NW of \snr\ in the 0.5--7.0 keV band, obtained using the {\scshape ciao} \emph{merge\_obs} script. The X-ray picture shows a bright rim of emission, presumably due to limb brightening at the forward shock of the SNR, as well as two thin arcs ahead of it (regions A and C) and faint diffuse emission behind and ahead of the shock (regions B, E, F, G, H, and I). In order to perform a like-to-like comparison, we convolved the X-ray image with a gaussian kernel of size 8$''$, and contrasted it to the Australian Telescope Compact Array 1.38 GHz radio image obtained by \citet{dickel_2001}, which has comparable resolution. These images of the NW region are notably different since the X-ray emission is concentrated in thin arcs and a main rim is clearly visible, while the radio picture is much more diffuse and extended. 

The two-color image shown in Figure \ref{fig:broad} (\emph{bottom left}) was created by combining the soft X-ray band (0.3--0.75 keV, in magenta) and the hard band (1.5--7 keV in blue). The thin arcs ahead of the shock appear to be dominated by hard X-rays, as is the eastern part of the main rim. On Figure \ref{fig:broad} (\emph{bottom right}) we show a two-color optical image of this region (S~{\scshape ii} in red, H$\alpha$ in cyan), created using observations with the 0.9m Curtis/Schmidt telescope at the Cerro Tololo Inter-American Observatory (CTIO) by \citet{smith_1997}. The stars in the frame have been removed and the image was median filtered over 9$''$. The entire main rim is bright in H$\alpha$ emission, and there is evidence of some diffuse emission in the region as well. Since the hard X-ray radiation corresponds well with the
Balmer-dominated emission (regions with bright H$\alpha$ and no S~{\scshape ii}), it appears clear that the main (hard X-ray) rim is associated with a non-radiative shock. Using Balmer-line spectra, \citet{ghavamian_1999} derives a shock velocity of $V_{\text{S}} = 650\pm120$ km s$^{-1}$ in this region. The thin hard X-ray arcs ahead of the shock, however, are not detected in the optical observations.

\begin{figure*}
\begin{center}
\includegraphics[width=0.40\textwidth]{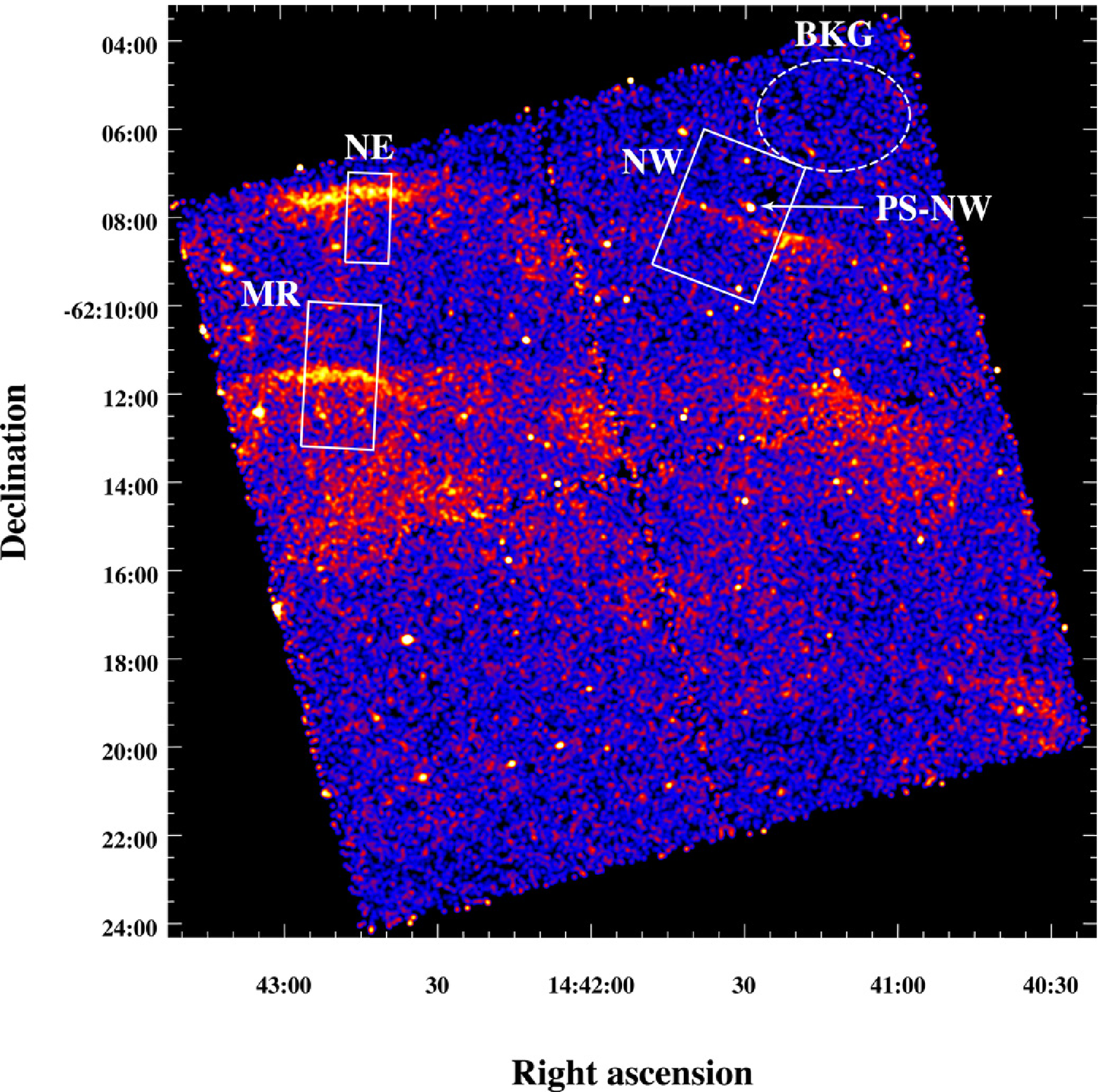}
\includegraphics[width=0.59\textwidth]{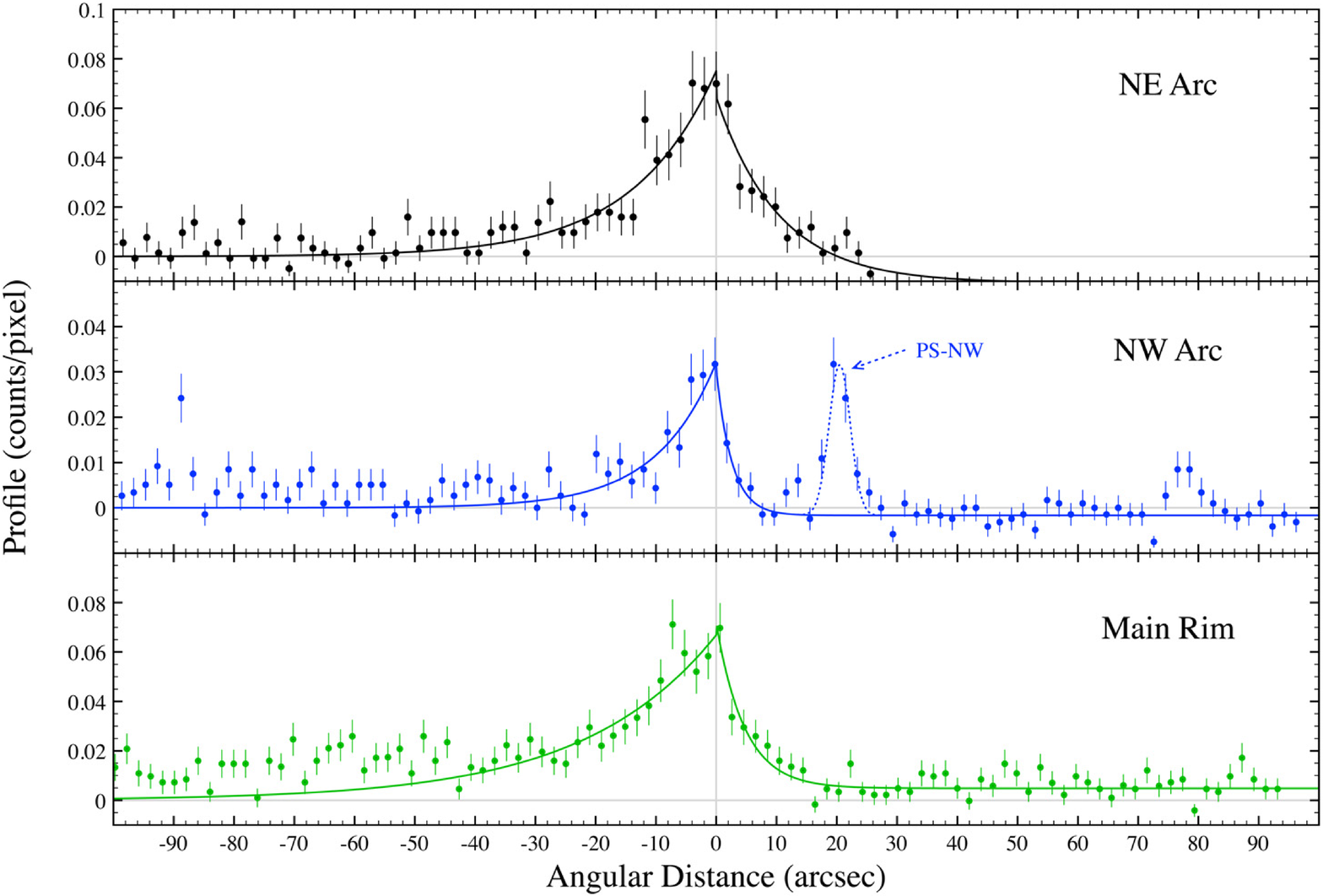}
\end{center}
\caption{\footnotesize \emph{Left panel:} Exposure corrected image in the 2.0--6.0 keV energy range, with pixel size 0$''$.492, smoothed with a gaussian with kernel $\sigma=10$ pixels, and using a square root intensity scale. Overlaid are shown the regions used for extracting emission profiles. \emph{Right panel:} Emission profiles extracted from the three prominent synchrotron filaments in the NW region of \snr. The filament widths obtained modeling the profiles using Equation 1 are presented in Table \ref{tab:B}. The profile of the NW arc shows an additional feature at angular distance $r\approx20$, which corresponds to a point source not believed to be a part of the SNR (identified as PS-NW). The best fit gaussian model fit to the profile of this point source was found to have full width half maximum of $4.1\pm0.6$, and is shown as a dotted line. This additional source is further discussed in \S2.}
\label{fig:profiles}
\end{figure*}

To constrain the characteristics of the non-thermal emitting material in the NW of \snr, we extracted spectra from 9 different regions, which are shown in white in Figure \ref{fig:broad} (\emph{top right}). 
Regions where thermal X-ray emission dominates will be discussed in a follow-up paper. The spectra (shown in Figure \ref{fig:spectra}) and corresponding weighted response files were obtained by using the {\scshape ciao} \emph{specextract} script. These spectra were fit using the {\scshape sherpa} modeling and fitting software package \citep{sherpa_2007}. We employed the \emph{chi2xspecvar} statistic, which uses $\chi^2$ with variance computed from data amplitudes in deriving the best-fit parameters. In all regions, the model that best describes the spectral characteristics is an absorbed non-thermal component combined with a non-equilibrium ionization collisional plasma, {\scshape nei}\footnote{{\scshape neivers} 2.0} \citep{borkowski_2001}. We adopt the Tuebingen-Boulder ISM absorption model \citep{wilms_2000}, and fix the absorbing column density to $N_{\text{H}}=0.6\times10^{22}$ cm$^{-2}$, which is the value obtained for a fit to all spectra combined. The abundances are also set to those from \citet{wilms_2000}. The non-thermal component is modeled using two different prescriptions, {\scshape powlaw1d} and {\scshape srcut} \citep{reynolds_1999}. While {\scshape powlaw1d} describes a power law photon distribution in X-ray energies, {\scshape srcut} models the synchrotron emission from an exponentially cut-off powerlaw electron distribution in a homogeneous magnetic field, which is in itself an exponentially cut-off powerlaw. The radio spectral index (at 1 GHz) for {\scshape srcut} is fixed at the value $\alpha=0.6$, as derived from radio observations \citep{green_2009}. Since the spectra are dominated by the non-thermal component, we set the electron temperature and ionization timescale of the {\scshape nei} component to the values obtained for the brightest thermal emission region (the eastern section of the main rim -- region D), $kT_{\text{e}}=0.56$ keV and $n_{\text{e}}t=10^{10}$s cm$^{-3}$, respectively. An additional component, a gaussian centered on 6.45 keV energy, was added to account for the Fe-K$\alpha$ emission detected in this region with \xmm\ and \suzaku\ \citep[][respectively]{williams_2011, yamaguchi_2011}. This Fe-K$\alpha$ emission is believed to be evidence of reverse shocked iron-rich ejecta located in unresolved clumps in this region. 
The parameters of the best-fit models for all extracted spectra are shown in Table \ref{tab:spectral}, where the uncertainties quoted are the 1$\sigma$ confidence limits. 

The energy flux in the 2--6 keV range is clearly dominated by non-thermal emission, and only in the main rim region (D) does the thermal flux surpass 5\% that of the non-thermal component. The cut-off energy of the spectrum (from {\scshape srcut}) varies between 0.1 and 1 keV, consistent with the overall spectral result from \citet{williams_2011} of $h\nu_{\text{roll-off}}\approx0.3$ keV. The value of $h\nu_{\text{roll-off}}$ depends strongly on the radio spectral index, and since there are no spatially resolved radio spectral studies of the region, the results of the fitting process should only be regarded as an approximate estimate. The power-law indices obtained using {\scshape powlaw1d} span values between 2.4--3.1. There are no clear differences in $\Gamma$ between the regions enclosing diffuse emission (i.e., B, E, F, G, H, and I) and those with rim-like regions features (A, C, and D). 

Figure \ref{fig:profiles} (\emph{left panel}) shows the exposure corrected \chandra\ image in the  2--6 keV band. To determine the filament emission widths of the rim-like features, we extracted the background subtracted emission profiles in the three regions shown. The profiles were obtained using strips of 4 pixels (1$''$.97) wide and with varying lengths ($a_{NE}=120$ pix, $a_{NW}=300$ pix, and $a_{MR}=200$ pix), as shown in Figure \ref{fig:profiles} (\emph{right panel}). These profiles were fit in {\scshape sherpa} using a model based on that of \citet{bamba_2005}, i.e.,
\begin{equation}
f(x) = \left\{
        \begin{array}{rlr}
        A\exp \left(-|\frac{x_0-x}{l_{\text{up}}}|\right) & x>x_0 \\
        A\exp \left(-|\frac{x_0-x}{l_{\text{down}}}|\right) & x<x_0,
        \end{array}
\right.\label{model}
\end{equation}
where $A$ and $x_0$ are the flux and position at the emission peak, and $l_{\text{up}}$ and $l_{\text{down}}$ are the characteristic scales of the emission profile in the upstream and downstream regions respectively. The results of the fits are included in Table 2, and the interpretation of these is described in Section 3. The profile of the NW arc, shown in the right panel of Figure \ref{fig:profiles}, includes an additional feature at angular distance $r\approx20''$, which corresponds to a point source not believed to be a part of the SNR. The best fit gaussian model fit to the profile of this point source was found to have full width at half maximum (FWHM) of $4.1\pm0.6$, and is shown as a dotted line. For reference we have marked the location of this source as "PS-NW" in the top right panel of Figure \ref{fig:broad}). \citet{allen_2004} derived a parameterization of the point spread function (PSF) of \chandra, and the expected blurring in the $2-6$ keV band, and at the off-axis angle of "PS-NW" ($\sim5'$), is consistent with that obtained from the gaussian fit to the profile. Hence, the profile of this additional source can be used as an approximate handle of the PSF of the instrument at the position of the NW arc. However, \citet{allen_2004} predict that a the approximate off-axis positions of the NE arc and main rim regions used in the profile analysis, $9'$ and $7'$ respectively, would correspond to larger PSFs ($13''-8''$). 

\section{Discussion}

\noindent Several authors \citep[e.g., ][]{vink_2003,parizot_2006} argue that the magnetic field strength of the post-shock region is closely linked to the width of X-ray synchrotron filaments in SNRs. \citet{vink_2003} propose that the width of the X-ray synchrotron-emitting filaments results from a combination of the velocity at which electrons are advected downstream from the shock, $v_{\text{adv}}$, and the timescale on which electrons lose energy through synchrotron radiation, $\tau_{\text{syn}}$. If one neglects electron diffusion, the stationary version of the transport equation at a parallel shock is $v_{\text{adv}}(\partial f/\partial x)=-f/\tau_{\text{syn}}$, with solution $f_{\text{e}}\propto \exp(-x/v_{\text{adv}}\tau_{\text{syn}})$ \citep{volk_1981}. The size of the advection region then is 
\begin{equation}
l_{\text{adv}}=v_{\text{adv}}\tau_{\text{syn}},
\end{equation}
 which can be connected with the downstream characteristic filament scale derived from observations above (i.e., fitting observations to the model described in Equation~1), through a factor $P$ so that $l_{\text{down}}=P l_{\text{adv}}$. This factor accounts for the shell geometry projection effect and a correction for the observed width resulting from a convolution of electron advection and diffusion. We have adopted $P=3.7$, as estimated by \citet{vink_2006}.

\begin{center}
\begin{deluxetable*}{c|ccc|ccc|cccc}
\tabletypesize{\footnotesize}
\tablecaption{\footnotesize{Profile Fit Parameters and Magnetic Field Estimates}\label{tab:B}}
\startdata
\toprule
\noalign{\smallskip}
\multirow{2}{*}{Region}&$l_{\text{down}}$&$l_{\text{up}}$&\multirow{2}{*}{$\chi^2/\text{(dof)}$}&$l_{\text{down}}$&$l_{\text{up}}$&$V_{\text{s}}$\tablenotemark{a}&$B_{\text{adv}}$\tablenotemark{b}&$B_{\text{diff,d}}$\tablenotemark{c}&$B_{\text{diff,u}}$\tablenotemark{c}&$B_{\text{joint}}$\tablenotemark{d}\\
&(arcsec)&(arcsec)&&(pc)&(pc)&(km s$^{-1}$)&($\mu$G)&($\mu$G)&($\mu$G)&($\mu$G)\\
\midrule
NE Arc&$14\pm2$&$10\pm3$&60 / (59)&$0.17\pm0.03$&$0.13\pm0.04$&$810\pm150$&27&300&140&110\\
NW Arc&$10\pm2$&$3\pm1$&94 / (74)&$0.12\pm0.02$&$0.03\pm0.02$&$810\pm150$&33&370&360&140\\
Main Rim&$22\pm2$&$5\pm1$&75 / (65)&$0.27\pm0.03$&$0.06\pm0.01$&$650\pm120$&17&250&280&80
\enddata 
\tablecomments{Both $l_{\text{down}}$ and $l_{\text{up}}$ are converted to pc using the assumed distance, $d=2.5$ keV.}
\tablenotetext{a}{Shock velocity derived from Balmer line profile \citep{ghavamian_1999} for the main rim region. The shock velocities quoted for the arcs ahead of the rim are linear projections of the Balmer-line profile velocity estimate.}
\tablenotetext{b}{Derived using the advection length method, using $l_{\text{down}}=P l_{\text{adv}}$, where $P=3.7$ is the projection factor, and Equation 5.}
\tablenotetext{c}{Estimates of the magnetic field obtained through the diffusion length method, using $l_{\text{down}}=P l_{\text{diff,d}}$ and $l_{\text{up}}=P l_{\text{diff,u}}$, where $P=3.7$ is the projection factor, and Equations 6 and 7.}
\tablenotetext{d}{Obtained using the $l_{\text{adv}}\approx l_{\text{diff,d}}$ condition (Equation 8).}

\end{deluxetable*}
\end{center}

Electrons of energy $E_{\text{e}}$ (in TeV) propagating through a magnetic field with strength $B$ (in $\mu$G), will emit synchrotron radiation with a characteristic photon energy peak at 
\begin{equation}
h\nu_{\text{peak}}\simeq \frac{B_{100}E_{\text{e}}^2}{520} \,{\rm  keV},
\end{equation}
where $B_{100}\equiv B/100$ \citep{pacholczyk_1970}. Once accelerated, the electron will radiate at energy $h\nu_{\text{peak}}$ for a time 

\begin{equation}
\tau_{\text{syn}} = (\text{1250 years}) E_{\text{e}}^{-1}  B_{100}^{-2},
\end{equation}
the synchrotron loss time. Combining Equations 2--4, for peak photon energies $h\nu_{\text{peak}}$ close to those of the logarithmic average of the observations in this work, 
3.5 keV, the magnetic field is 
\begin{equation}
B_{\text{adv}} 
\approx (\text{83 }\mu\text{G}) \left(\frac{l_{\text{adv}}}{0.01\text{ pc}}\right)^{-2/3} \left(\frac{V_{\text{S}}}{1000\text{ km s$^{-1}$}}\right)^{2/3},
\end{equation}
using $v_{\text{adv}}=V_{\text{S}}/\chi_S$, where $\chi_S$ is the shock compression ratio. We have adopted $\chi_S=4$, which is that expected for a strong shock unmodified by particle acceleration, although the material is expected to be more compressible in reality due to diffusive shock acceleration \citep[][and references therein]{castro_2011}. The inferred magnetic field $B_{\text{adv}}$ should therefore have a scaling dependence on the compression ratio as $(\chi_S/4)^{-2/3}$.

As mentioned above, the velocity of the shock at the main NW rim of \snr\ was estimated to be $V_{\text{S}}=650\pm120$ km s$^{-1}$ using Balmer line profiles by \citet{ghavamian_1999}. Since we have calculated the smaller non-thermal arcs ahead of the main rim to be approximately 5$'$ beyond it, using a distance of $d=2.5$ kpc, we extrapolate the velocity of the shock at their position to be approximately $810\pm150$ km s$^{-1}$. 
Table \ref{tab:B} lists the derived magnetic fields corresponding to the filament widths calculated in Section 2 using Equation 5.

An alternative approach is that proposed by \citet{bamba_2004}, and \citet{volk_2005}, which assumes that the filament widths downstream and upstream correspond to the diffusion length scales, $l_{\text{diff}}=D/v_{\text{diff}}$, where $D$ is the diffusion coefficient, and $v_{\text{diff}}$ is the bulk flow relative to the shock frame. In the "Bohm limit", the smallest possible diffusion coefficient for isotropic turbulence is assumed, where the electron mean free path equals the Larmor or gyroradius and hence $D=cE_{\text{e}}/3eB$ \citep{parizot_2006}. One can then derive the upstream and downstream magnetic fields as a function of diffusion lengths and the shock velocity, i.e.,
\begin{equation}
B_{\text{diff,d}} \approx (\text{700 }\mu\text{G}) \left(\frac{l_{\text{diff,d}}}{0.01\text{ pc}}\right)^{-2/3} \left(\frac{V_{\text{S}}}{1000\text{ km s$^{-1}$}}\right)^{-2/3},
\end{equation}

\noindent for the downstream field (using $l_{\text{down}}=P l_{\text{diff,d}}$ and $v_{\text{diff,d}}=V_{\text{S}}/4$), and 

\begin{equation}
B_{\text{diff,u}} \approx (\text{280 }\mu\text{G}) \left(\frac{l_{\text{diff,u}}}{0.01\text{ pc}}\right)^{-2/3} \left(\frac{V_{\text{S}}}{1000\text{ km s$^{-1}$}}\right)^{-2/3},
\end{equation}

\noindent for the upstream region, using (using $l_{\text{up}}=P l_{\text{diff,u}}$ and $v_{\text{diff,u}}=V_{\text{S}}$). The values derived from the observations of the NW region of \snr\ are estimated to lie between 300 and 400 $\mu$G, and are included also in Table 2. While in other cases the two different methods yield similar estimates of the magnetic field \citep[e.g.][]{vink_2005,ballet_2006}, \citet{vink_2006} found the values derived for the NE of \snr\ to be inconsistent. This is also the case for the magnetic field estimates gathered in this work. It is very likely that this discrepancy arises from the shock velocity at these filaments being much higher than the value derived through Balmer filament profiles. If one were to assume that the shock is in its adiabatic evolution phase \citep{castro_2011}, and use the estimated distance $d=2.5$ kpc, and age $t=1830$ years, the derived expansion velocity at the position of these filaments would be $\sim3000-4000$ km s$^{-1}$, which would yield more consistent results between the two methods. 

A third method to connect filament width with magnetic field is based on using the condition where the advection length is approximately equal to the downstream diffusion length of electrons, $l_{\text{adv}}\approx l_{\text{diff,d}}$. This provides a shock velocity independent method for estimating the magnetic field strength. The condition itself, \citet{vink_2012} argues, holds for electrons close to the maximum energy, which is an appropriate assumption in the case of \snr. The expression for the magnetic field in this case is then,
\begin{equation}
B_{\text{joint}} \approx (\text{240 }\mu\text{G}) \left(\frac{l_{\text{adv}}}{0.01\text{ pc}}\right)^{-2/3}.
\end{equation}
The estimates derived using this prescription are shown in Table 2, and range between 80 and 140 $\mu$G. As discussed in \S2, the PSF of \chandra\ in the $2-6$ keV band is comparable or larger than the length scales derived from the fits to the emission profiles, and hence the filament widths obtained are upper limits. Additionally, the Bohm diffusion coefficient employed represents the lower limit for particles propagating in these magnetized environments. Hence, the magnetic fields derived and shown in Table 2 are lower-limits on these magnitudes. If one considers that typical magnetic fields in the ISM are approximately 1--10 $\mu$G, the values derived from the observations clearly indicate very significant magnetic field amplification factors of $\sim10-100$.

Using the $l_{\text{adv}}\approx l_{\text{diff}}$ condition one can also calculate the shock speed based on the shape of the exponential cut-off of the photon spectrum. \citet{vink_2006} derives the expression to be

\begin{equation}
V_{\text{S}} \approx (\text{2650 km s}^{-1}) \left(\frac{h\nu_{\text{peak}}}{1\text{ keV}}\right)^{1/2},
\end{equation}
where the compression ratio has been set to $\chi_S=4$. We can hence estimate, taking the peak photon energy derived from the spectra (and shown in Table 1) $h\nu_{\text{peak}}\approx0.1-1.0$ keV, that the shock velocity in the region is approximately 800--2700 km s$^{-1}$. Furthermore, a rough estimated range for the maximum electron energy between 10 and 20 TeV can be obtained combining Equation 3 with the estimates for the cut-off energy (0.1--1.0 keV) from the spectral fits, and the values for $B_{\text{joint}}$ shown in Table \ref{tab:B}. 
 
The assumption that the arc features found ahead of the main shock are physically in the same plane as the brighter rim is an oversimplification, and the shock velocities derived hence are very conservative lower limits. It appears clear that the estimates of the synchrotron filament widths obtained in this work indicate higher shock velocities for these non-thermal regions than those derived for the Balmer filaments. It is possible that the shock velocity obtained through Balmer line profiles is an underestimate due to shock modification resulting from the particle acceleration process itself \citep[e.g.][]{castro_2011}.
 
The magnitudes of the magnetic fields estimated in similar studies of SNRs Cas A, Kepler, Tycho and SN 1006, lie in the range 30 $\mu$G to 300 $\mu$G \citep{vink_2006b}, and the magnetic field strengths derived in this work are approximately within that range. However, the nature of \snr\ appears to be somewhat different than that of these other X-ray synchrotron emitting sources, since the overall morphology of its non-thermal X-ray emission is much more irregular. \citet{williams_2011} argue that the shock of \snr\ propagated through a low density bubble and has only recently started interacting asymmetrically with a denser shell of material from a late-phase progenitor wind,  giving rise to its peculiar shape. In addition to its morphological differences, the shock velocities estimated for Cas A, Kepler, Tycho and SN 1006, are in the $3000-5400$ km s$^{-1}$ range. In contrast, the shock velocities obtained through optical studies of the Northern, Northwestern, Eastern, and Southwestern regions of \snr\ are much smaller, varying between 300 and 900 km s$^{-1}$ \citep{ghavamian_1999}. As we argued before, it is likely that the velocities of the synchrotron emitting shocks in \snr\ are higher than those inferred from Balmer-line observations, and recent proper motion studies based on optical observations of the E rim of \snr\ estimate the shock velocity in that region to vary between 700 and 2200 km s$^{-1}$ \citep{helder_2013}. In any case, it is safe to conclude that \snr\ stands as peculiar example in the class of SNRs with X-ray synchrotron filaments due to its irregular morphology and low shock velocities.

\acknowledgments

The authors thank Parviz Ghavamian for some important discussions and insight. DC acknowledges support for this work provided by the Chandra GO grant GO3-14080, as well as, the National Aeronautics and Space Administration through the Smithsonian Astrophysical Observatory contract SV3-73016 to MIT for Support of the Chandra X-Ray Center, which is operated by the Smithsonian Astrophysical Observatory for and on behalf of the National Aeronautics Space Administration under contract NAS8-03060. PC acknowledges support from National Aeronautics Space Administration under contract NAS8-03060.

\bibliography{ms-2col}

\end{document}